\documentclass[11pt,twoside]{article}
\usepackage[utf8]{inputenc}
\usepackage{color}
\usepackage{amssymb}
\usepackage{amsfonts}
\usepackage{amsmath}
\usepackage[english]{babel}
\usepackage[pdftex]{graphicx}
\usepackage{fancyhdr}
\usepackage[normalem]{ulem}

\newcommand{\beq}{\begin{equation}}
\newcommand{\eeq}{\end{equation}}
\newcommand{\hone}{h_1}

\newcommand{\hpm}{h^{\pm}}
\newcommand{\mhpm}{m_{h^{\pm}}}
\newcommand{\met}{E\!\!\!\!/_T}

\begin{document}

\fancypagestyle{plain}{%
\fancyhf{}%
\fancyhead[LO, RE]{XXXVIII International Symposium on Physics in Collision, \\ Bogot\'a, Colombia, 11-15 September 2018}}

\fancyhead{}%
\fancyhead[LO, RE]{XXXVIII International Symposium on Physics in Collision, \\ Bogot\'a, Colombia, 11-15 September 2018}

\title{Neutral and Charged Higgs boson phenomenology \\ at the LHeC and FCC-eh}
\author{\uline{Siba Prasad Das}$\thanks{e-mail: sp.das@uniandes.edu.co}$, 
Marek Nowakowski$\thanks{e-mail: nowakows@uniandes.edu.co}$,\\
Department of Physics, Faculty of Science,
Universidad de los Andes, \\ Apartado Aereo 4976-12340, Carrera 1 18A-10, Bloque Ip, 
Bogota -- Colombia\\
J. Hern\'andez-S\'anchez$\thanks{e-mail: jaime.hernandez@correo.buap.mx}$\\ 
Facultad de Ciencias de la Electr\'onica, Benem\'erita Universidad Aut\'onoma de Puebla,\\ 
Apdo. Postal 542, C.P. 72570 Puebla, Puebla, M\'exico\\
Stefano Moretti$\thanks{e-mail: s.moretti@soton.ac.uk}$\\ 
School of Physics and Astronomy, University of Southampton, \\Highfield, Southampton SO17 1BJ, United Kingdom.\\
Alfonso Rosado$\thanks{e-mail: rosado@ifuap.buap.mx}$\\
Instituto de F\'{\i}sica, Benem\'erita Universidad Aut\'onoma de Puebla, \\ Apdo. Postal J-48, C.P. 72570 Puebla, Puebla, M\'exico\\
}
\date{\today}
\maketitle

\begin{abstract}
  We have analyzed the prospects of observing the  lightest CP-even neutral Higgs boson ($h_1$) 
  via decays into $b \bar b$ pairs in the neutral and charged current production
  processes $e^- q \to e^- h_1 q$ and $e^- q\to \nu_e h_1 q'$, respectively,  at the planned 
  Large Hadron electron Collider (LHeC), with an $e^-$ beam energy of  60 GeV and a $p$ beam 
  energy of 7 TeV. We also focus on observing a relatively light charged Higgs boson ($h^{-}$) 
  via its production  mode $e^- b \to \nu_e h^{-} b$   
  followed by the decays  $h^{-} \to s \bar c + s \bar u$ at the upcoming Future Circular Collider 
  in hadron-electron  mode (FCC-eh) with $\sqrt s \approx 3.5$ TeV. We have performed our analysis 
  in the framework of the Next-to-Minimal Supersymmetric Standard Model (NMSSM) wherein the intermediate 
  Higgs boson ($h_2$) is Standard Model (SM)-like. We have considered constraints from Dark Matter (DM), 
  superparticle and the Higgs boson data. In both analyses, we have carried out  signal and background 
  computations with selections optimized to the model at hand  and found that in both cases it is possible to get large
  significances for low mass Higgs bosons. 

\end{abstract}

\section{Introduction}

Since the discovery of a Higgs boson, with a mass of 125 GeV, at the Large Hadron Collider (LHC), by the 
ATLAS and CMS experiments,  the SM with  spontaneous Electro-Weak Symmetry Breaking (EWSB) 
is apparently well established. The SM contains one doublet of Higgs isospin. Nonetheless, many models with 
enlarged Higgs sectors still survive LHC SM-like Higgs data. In fact, any deviations from SM predictions 
would be a hint in favor of new physics in Nature \cite{Ellis:2017upx}. Whereas several new physics 
scenarios exist that cannot only comply with the aforementioned LHC results (as well as explain other experimental observations that cannot be accounted for in the SM, such as neutrino 
 and DM data) but also provide motivated theoretical frameworks (e.g., solving the hierarchy 
problem of the SM), it is fair to say that Supersymmetry (SUSY) is one of the most appealing ones.

However, it is very well known that SUSY in its minimal incarnation, called the MSSM, has several flaws. On the theoretical side, it suffers from the $\mu$-problem, as this parameter 
(effectively mixing the SUSY counterparts of Higgs states) ought to be below the TeV scale in order to enable successful EWSB, yet in the MSSM it can really naturally be only zero or close to the Planck mass. On the experimental side, its allowed parameter space is being more and more constrained from nil searches for new Higgs bosons or SUSY states. Both problems are remedied in the so called NMSSM \cite{nmssmue,nmssmmd}, wherein the Vacuum Expectation Value (VEV) of an additional Higgs singlet state  can generate the $\mu $-term at the required  scale and its SUSY counterpart can alleviate experimental bounds as it can act as a new DM state simultaneously altering  SUSY cascade signals and the cosmological relic density. Just like in the MSSM, also the NMSSM has a pair of charged Higgs bosons ($h^\pm$) in its spectrum plus the possibility of a neutral CP-even Higgs boson lighter than the one
discovered. In fact, a myriad of other non-minimal SUSY scenarios also have these states  \cite{Khalil:2132388}.

Non-standard neutral  as well as charged Higgs bosons have been the focus of many analyses  at 
the LHC. These searches are generally performed model-independently and then 
interpreted in specific scenarios, like the MSSM or NMSSM. Both Higgs states are normally searched for via flavor diagonal decays.
More recently, the case for studying  the (non-diagonal) flavor
decay $ h^- \to b \bar{c}$ has also vigorously been made in a variety of new physics scenarios thus 
encouraging the LHC experimental groups to look for this signal too \cite{Akeroyd:2014cma}. 

At CERN, the future Large Hadron electron Collider (LHeC) and electron-proton Future Circular Collider (FCC-eh), 
with center-of-mass energies of 1.3 TeV and 3.5 TeV, respectively ~\cite{AbelleiraFernandez:2012cc,mellado,klein}, 
offer good prospects as Higgs boson factories, wherein one could elucidate the nature of the couplings of Higgs bosons to fermions, 
especially the $ b \bar{b} $ one, which is difficult to establish at the LHC, but also, e.g., of charged Higgs bosons to generic fermions. 
Given these encouraging prospects, we specifically analyze here the prospects of observing relatively light neutral and 
charged Higgs bosons of the NMSSM decaying via $b\bar b$ and  $ s \bar c + s \bar u$ modes, respectively. 

\section{NMSSM}

We just mention here the relevant parts of the NMSSM. The superpotential is described as

\beq
W_\mathrm{Higgs} = (\mu + \lambda \widehat{S})\,\widehat{H}_u \cdot
\widehat{H}_d + \xi_F \widehat{S} + \frac{1}{2} \mu' \widehat{S}^2 +
\frac{\kappa}{3} \widehat{S}^3,
\eeq

\beq
W_\mathrm{Yukawa} = h_u\, \widehat{Q} \cdot \widehat{H}_u\;
\widehat{U}^c_R + h_d\, \widehat{H}_d \cdot \widehat{Q}\;
\widehat{D}^c_R + h_e\, \widehat{H}_d \cdot \widehat{L}\;
\widehat{E}_R^c,
\eeq

\beq
W_\mathrm{} = \lambda \widehat{S}\,\widehat{H}_u \cdot
\widehat{H}_d + \frac{\kappa}{3} \widehat{S}^3.
\eeq

The effective $\mu$-term is $\mu_\mathrm{eff} = \lambda s$, where $s$ is the singlet VEV. Further,
 we invoke a simpler scenario in our analysis, namely,the $\mathbb{Z}_3$ symmetric version, wherein 
one can set  $\mu = \mu' = \xi_F = 0$ and $m_3^2 = m_{S}'^2 = \xi_S = 0$. In this model 
we have three neutral scalar fields and they mix to form three mass eigenstates, generally describes 
as $h_1$, $h_2$ and $h_3$ (in increasing order of mass), where $h_2$ is considered as the SM-Higgs one  in our analysis. 

\section{Numerical Analysis}

Details of the tools used in our numerical analysis can be found in Refs.~\cite{Das:2016eob} and  \cite{Das:2018vuk}, including definition of the kinematic variables below.

\subsection{Neutral Higgs: $e^- p \to e^- \hone q$ and $e^- p\to \nu_e \hone q$ at the LHeC}

\begin{figure}[t!]
\begin{center}
\includegraphics[width=0.49\textwidth]{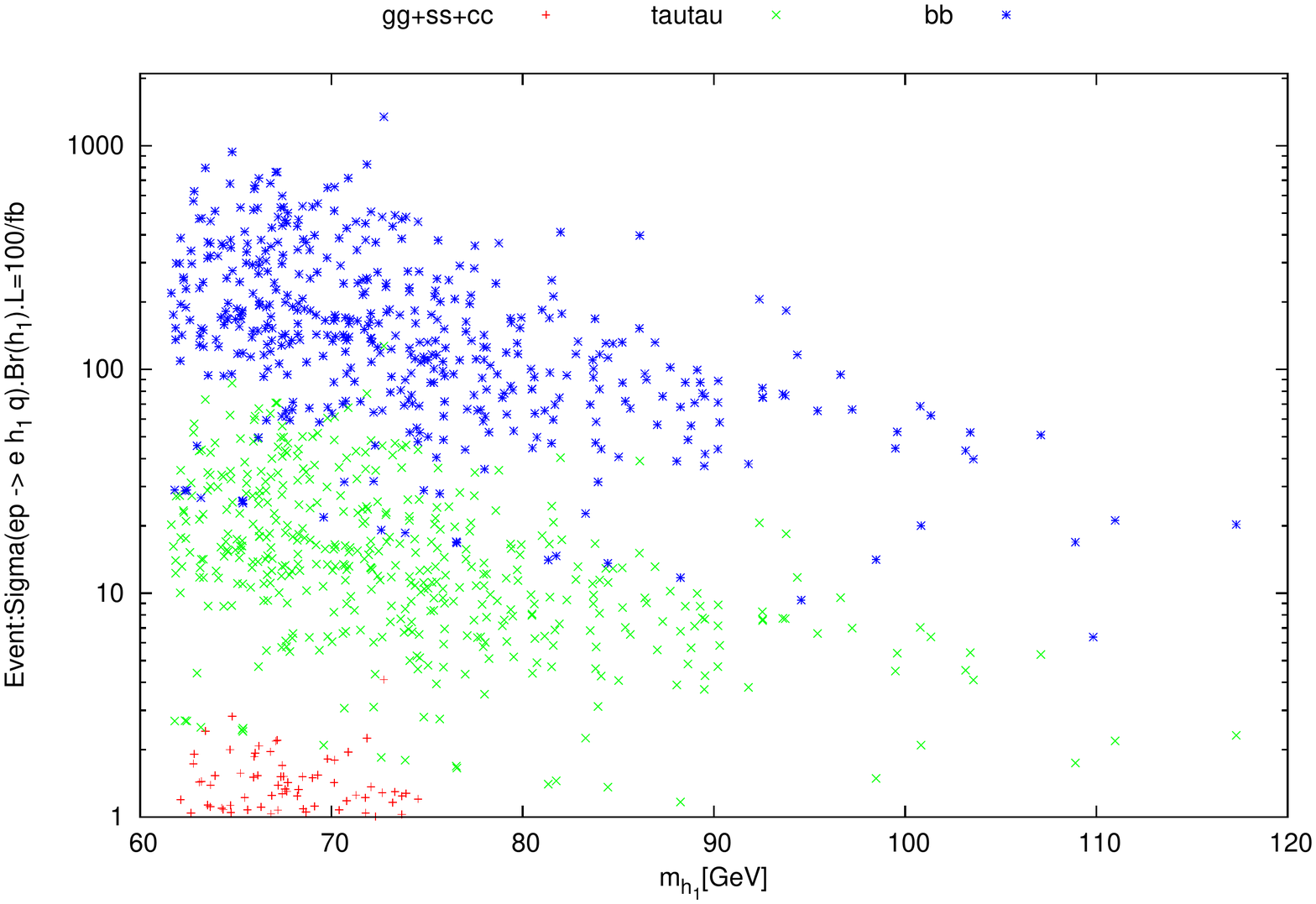}
\includegraphics[width=0.49\textwidth]{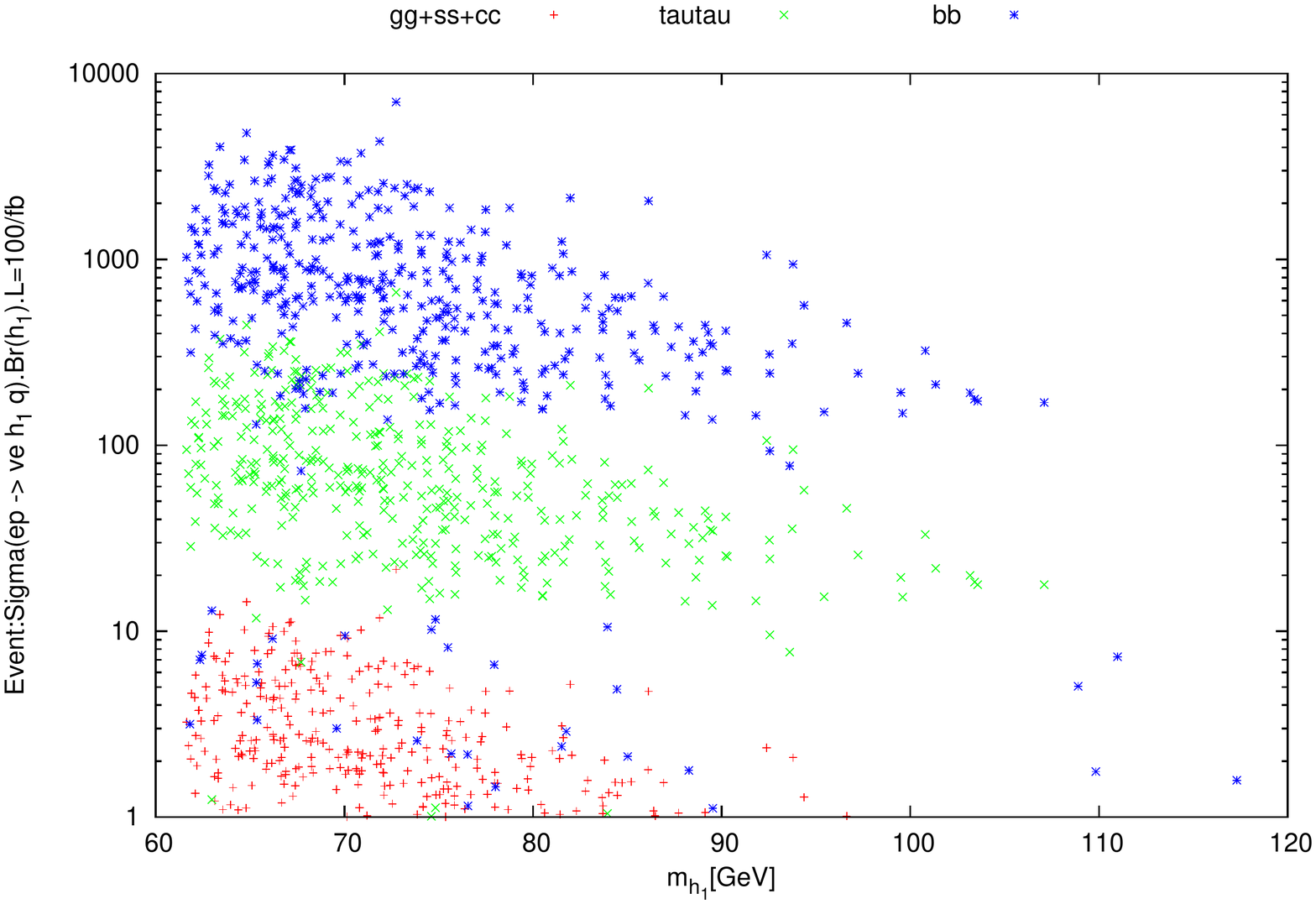}
\vskip-0.8cm
\caption{Event rates for $e^- q \to e^- \hone q$ (left) and $e^- q\to e^- \nu \hone q$ (right) at the LHeC.}
\end{center}
\label{LHeC-rates}
\end{figure}

Starting from the inclusive rates in Fig.~1 and after seeing  various differential distributions (see details in \cite{Das:2016eob}), we applied various selection cuts  
to isolate the Signal (S) from the Background (B) and performed  some simple optimization 
to enhance the significances. In particular, we varied the following parameters over the ranges (min,max,step): $\eta_{l}$$(1.0,2.5,0.1)$, 
$\eta_{l}$$(-2.5,-1.0,0.1)$, $\Delta {\eta_{jl}}$$(0.0,1.5,0.1)$, 
$\Delta {\eta_{jl}}$ $(-6.0,-3.0,0.1)$, $m_{\phi j}$(80,180,10) GeV, $H^j_T$(70,140,10) GeV and $|\vec H^j_{T}|$(30,60, 10) GeV.  
In Table ~\ref{nmssm1}, we show some Benchmark Points (BPs) and corresponding 
optimized significances. Signal extraction can be achievable at nearly 3$\sigma$   with a low luminosity collider option while full discovery relies on higher data samples\footnote{This mirrors the results of \cite{Das:2015kea}, albeit in another model.}.

\begin{table}[t!]
\centering
\caption{S and B rates together with significances (${\cal S}$) for the $h_1\to b\bar b$ signals at the LHeC as function of the optimized cuts obtained for 100 fb$^{-1}$(1 ab$^{-1}$).}\vspace*{0.1cm}
\hspace*{-0.75cm}
\scalebox{0.65}{
\begin{tabular}{||c||c|c|c|c||c|c|c|c||}
\hline
BP, $m_{h_1}$ (GeV)& $\eta_{l}$, $\Delta {\eta_{jl}}$, $m_{\phi j}$ (GeV), $H_T$ (GeV), $\vec H_T$ (GeV)&S&B&${\cal S}$&$\eta_{l}$, $\Delta {\eta_{jl}}$, $m_{\phi j}$ (GeV),$H_T$ (GeV), $\vec H_T$ (GeV) &S&B&${\cal S}$\\
\hline
$e1$,~63.59 &$(1.0,-1.0),(0.0,-4.3),180,130,60$ &4.9 &162.3 &0.38(1.2)&$(1.6,-2.5),(0.3,-6.0),100,140,30$&12.8&412.7&$0.63(1.99)$\\
$e2$,~70.59 &$(1.0,-1.0),(0.0,-3.0),180,140,60$ &2.7 &1.3 &2.36(7.5)&$(1.1,-2.5),(0.2,-5.7),90,90,30$&10.1&1295.3&0.28(0.89)\\
$e3$,~75.29 &$(1.0,-2.5),(0.4,-3.4),180,140,60$ &3.1 &1.5 &2.53(8.0)&$(1.0,-2.1),(0.4,-6.0),120,110,30$&11.6&565.2&0.49(1.54)\\
$e4$,~82.24 &$(1.0,-1.4),(0.0,-3.4),180,140,60$ &1.6 &0.6 &$2.09(6.6)$&$(1.0,-2.1),(0.1,-3.4),110,140,30$&4.1&154.1&0.32(1.01)\\
$e5$,~88.07 &$(1.0,-1.8),(0.0,-3.0),180,140,60$ &1.3 &2.4 &0.85(2.7)&$(1.3,-2.1),(0.1,-5.9),150,140,30$&4.8&340.0&0.26(0.82)\\
\hline
\end{tabular}
\label{nmssm1}
}
\end{table}

\subsection{Charged Higgs: $e b(\bar b) \to e \hpm b(\bar b)$ at the FCC-eh}

\begin{figure}[t!]
\begin{center}
\includegraphics[width=0.49\textwidth]{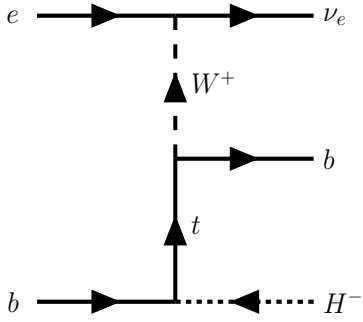}
\includegraphics[width=0.49\textwidth]{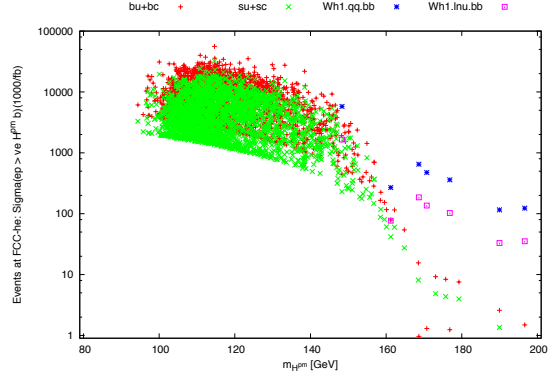}
\vskip-0.95cm
\caption{Feynman diagram (left) and event rates (right) for $e b (\bar b) \to e \hpm b(\bar b)$ at the FCC-eh.}
\label{feynevt}
\end{center}
\end{figure}

\begin{figure}[t!]
\begin{center}
\includegraphics[width=0.49\textwidth]{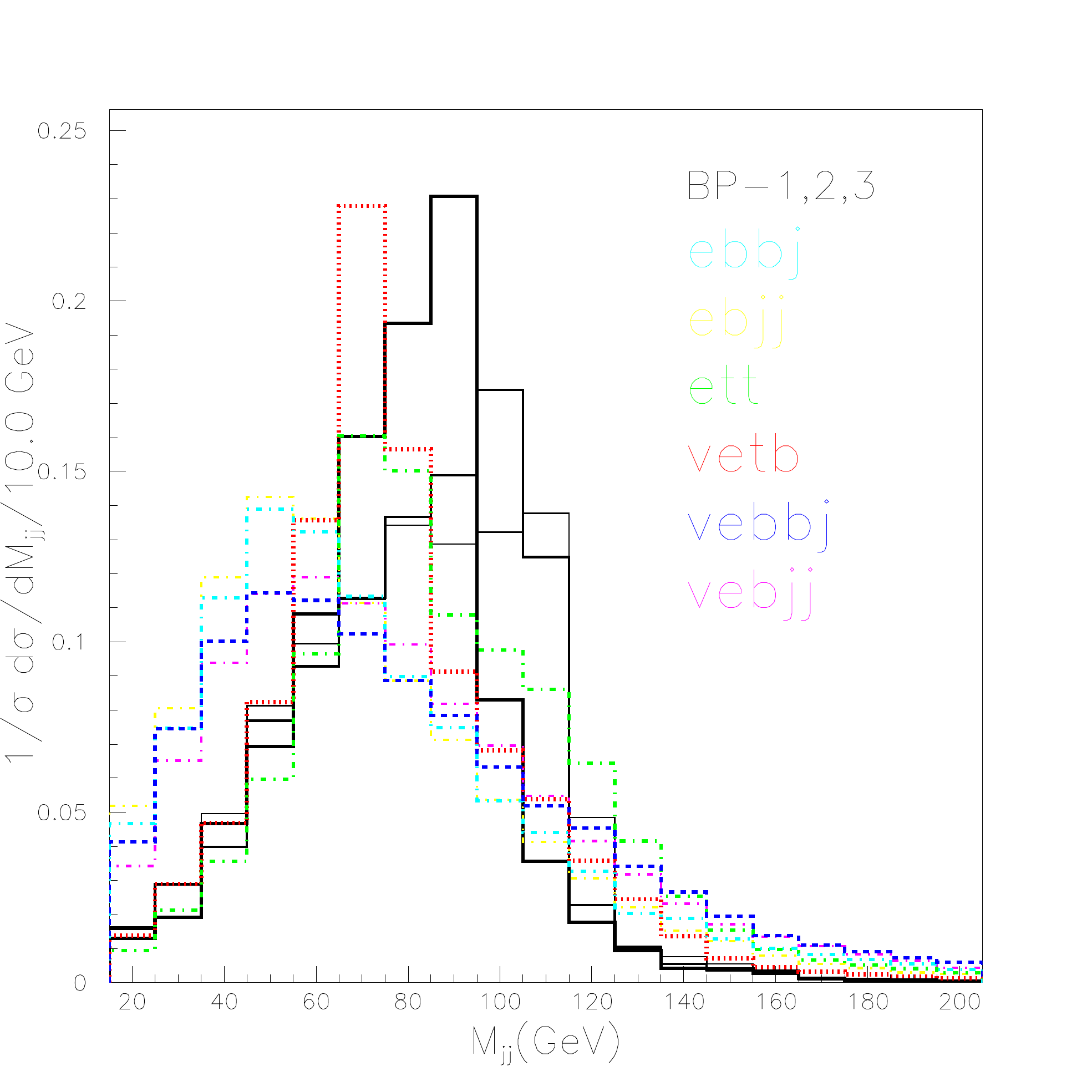}
\includegraphics[width=0.49\textwidth]{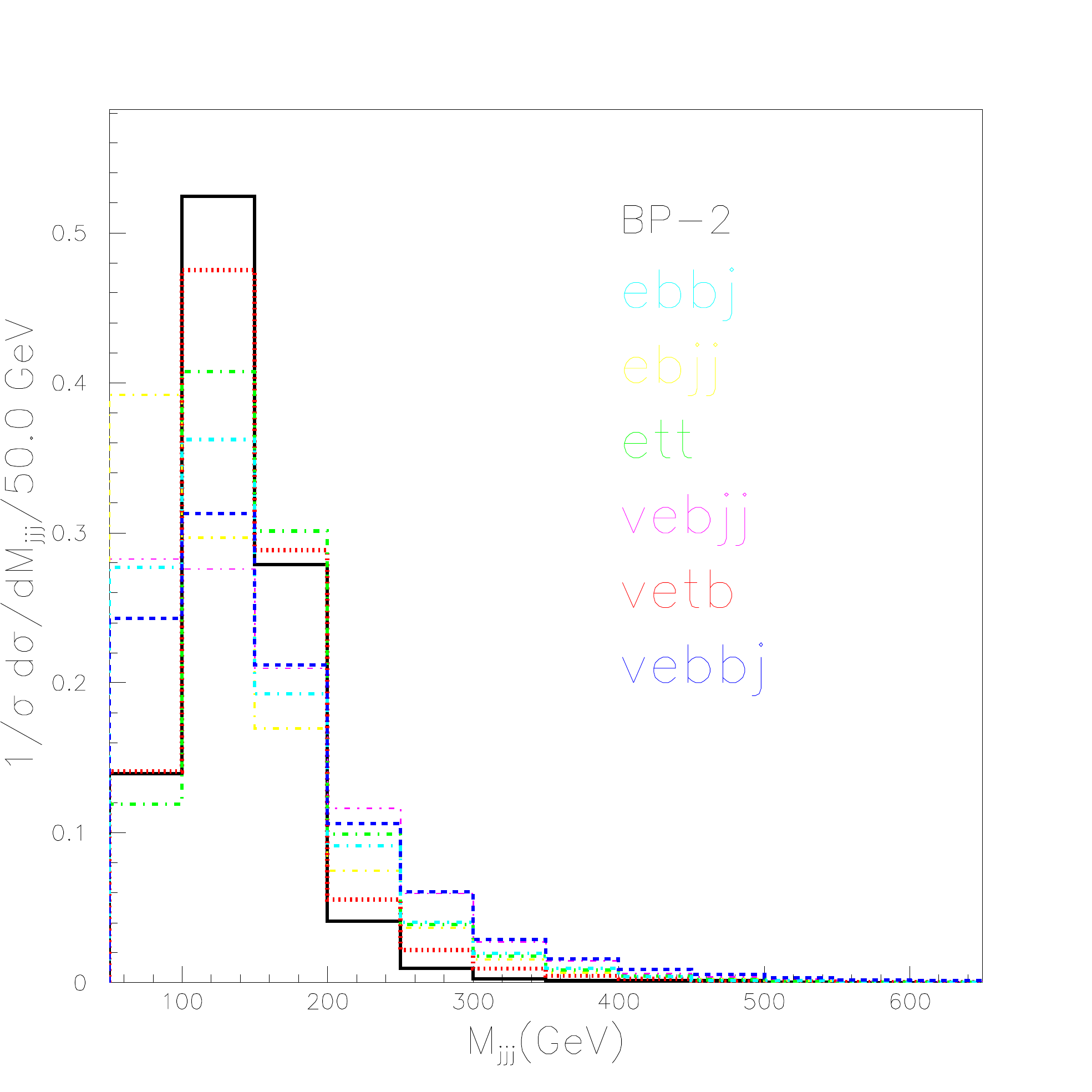}
\vspace*{-0.45cm}
\caption{Reconstructed di-jet (left) and final state (right) mass for $h^-\to s\bar c+s\bar u$.}
\label{masshpm}
\end{center}
\end{figure}

We show the Feynman diagram of these processes in the left-panel of Fig.~\ref{feynevt}. The inclusive event rate  
at FCC-eh energies is  instead shown in the right-panel of Fig.~\ref{feynevt}. From the whole allowed parameter space, 
we selected three BPs where the number of signal events is substantial. Like
in the previous case, upon seeing various kinematical distributions, we initially applied a simple cuts and count method, 
which  lead to  small significances. To establish the mass of the charged Higgs boson,  
we show in the left-panel of Fig.~\ref{masshpm} the di-jet invariant mass ($m_{jj}$). By combining 
together the charged Higgs boson candidate jets and a forward $b$-tagged jet, thus constructing the final state mass ($m_{jjb}$),  some interesting features  
appear against the  SM backgrounds (right panel of Fig.~\ref{masshpm}).  

We thus performed again a multi-dimensional grid optimization as follows: 
$\met$$(20.0,40.0,5)$ GeV, $H_T$ $(95.0,110.0,5)$ GeV,  $|\vec H_T|$$(20.0,40.0,5)$ GeV,  $R_M$$(2.5,3.5,0.025)$,  
the upper value of $m_{jj}$($\mhpm$, $\mhpm$+10.0,2.5) GeV,  
the lower value  of $m_{jj}$($\mhpm$--25.0 GeV, $\mhpm$--15.0,2.5) GeV,
the upper value of cos($\phi_{jj}$)$(0.45,0.55,0.01)$ and 
the upper value of $\Delta R$($\eta_{jj}, \phi_{jj}$)$(2.1,3.5, 0.1)$. For each of the generated  
combinations we estimated 
the number of S  and B events plus  the significances ${\cal S}$. 
As shown in Table ~\ref{tab:optlep}, there indeed exist combinations for which both evidence and (near) 
discovery of our $\hpm$ signals can be established, albeit only at 1 ab$^{-1}$ of luminosity.

\begin{table}[t!]
\centering
\caption{S and B rates together with significances (${\cal S}$) for the $h^-\to s\bar c+s\bar u$ signals at the FCC-eh as function of the optimized cuts obtained for 100 fb$^{-1}$(1 ab$^{-1}$).}\vspace*{0.1cm}
\scalebox{0.7}{
\begin{tabular}{||c||c|c|c|c|c|c|c|c|c|c|c||}
\hline
BP, $\mhpm$ (GeV) &$\met$&$H_T$&$\vec H_T$&$M_{jj}$$<$&$M_{jj}$$>$&$R_M$$<$& cos($\phi_{jj}$)$<$ & $\Delta R$($\phi_{jj}$)$<$&S&B&${\cal S}$\\
\hline
\hline
& 20.00 &105.00 & 20.00 &98.40 & 80.90 &  2.50 &  0.52 &  2.10 & 35.1& 9055.5 & 0.37(1.18)\\
BP1, 98.4&  20.00 &100.00 & 20.00 & 98.40 & 75.90 & 2.50 &  0.52 &  3.10 & 49.4 & 19714.8 & 0.35(1.12)\\
& 20.00 &105.00 & 20.00 & 103.40 & 73.40 &  2.50 &  0.52 &  3.10 & 55.8 & 27072.3 & 0.34(1.08)\\
\hline
\hline
& 20.00 &110.00 & 20.00 &114.60 & 89.60 &  2.50 &  0.54 &  2.90&145.3 & 11027.8 & 1.38(4.43)\\
BP2, 114.6& 20.00 &110.00 & 20.00 &114.60 & 99.60 &  2.50 &  0.54 &  2.20 & 86.6& 4890.5&1.24(3.96) \\
& 30.00 &110.00 & 30.00 &114.60 & 97.10 &  2.50 &  0.45 &  2.10 & 74.7 & 5005.8 & 1.05(3.38)\\
\hline
\hline
& 20.00 & 95.00 & 20.00 & 121.30 & 96.30 &  2.50 &  0.45 & 2.80 & 61.5& 8327.2 & 0.67(2.16)\\
BP3, 121.3& 20.00 &110.00 & 20.00 & 121.30 & 96.30  &  2.50 & 0.45 &  2.20 & 54.7& 7040.8  & 0.65(2.08)\\
&20.00 &100.00 & 20.00 & 121.30 &103.80 &  2.50 &  0.45 &  2.60 & 44.4& 5234.5& 0.61(1.96)\\
\hline
\end{tabular}
}
\label{tab:optlep}
\end{table}

\section{Conclusions}

In the NMSSM, a $h_1\to b\bar b$ signal (with $h_1$ lighter than the SM-like state seen at the LHC) can be
discovered at the LHeC with up to approximately 3(8)$\sigma$ significance for 100fb$^{-1}$(1 ab$^{-1}$) of  luminosity. Further, at the 
FCC-eh with 100 fb$^{-1}$(1 ab$^{-1}$) of integrated luminosity, charged Higgs signals $h^-\to s\bar c+s\bar u$  could achieve 
significances up to $\approx$$4.4(2.2)$$\sigma$ for  $m_{h^\pm}=114(121)$ GeV.\\[0.25cm]
\noindent
\underline{\bf Acknowledgments}~SPD acknowledges the High Performance Computing (HPC) facility at Universidad de los Andes. SM is supported in
part by the NExT Institute, STFC Consolidated Grant ST/L000296/1 and H2020-MSCA-RISE-2014 grant no. 645722
(NonMinimalHiggs). AR and JH-S are supported by SNI-CONACYT (M\'exico), VIEP-BUAP and  PRODEP-SEP (M\'exico)
under the grant: ``Red Tem\'atica: F\'{\i}sica del Higgs y del Sabor".

\end{document}